 \def\Title{Hardy's Second Axiom is insufficiently general}
 \def\arXiv{quant-ph/0302158}%{quant-ph/}
 \def\Abstract{%
Hardy (quant-ph/0101012) conjectures in his Axiom 2 that $K=K(N)$, and that in classical
probability $K=N$, while in quantum mechanics $K=N^2$. We offer an example in classical
probability for which $K=NV$, $V$ the number of independent complete variables; with
$N=V$ this classical example satisfies the purported quantal relation $K=N^2$.
}%

\documentclass[aps,prb,nobibnotes,nofootinbib,onecolumn,twoside,noshowpacs,fleqn,draft]{revtex4}
 \usepackage[tbtags,sumlimits]{amsmath}
 \usepackage{amssymb}
 \usepackage{xspace}

 \newcommand{\ie}{i.e., }

 \newcommand{\RefCites}[1]{Refs.~\onlinecite{#1}}

 \newcommand{\set}[1]{\ensuremath{{\left\{\,#1\,\right\}}}}%
 \newcommand{\K}{\ensuremath{\text{\textsf{K}}}\xspace}
 \newcommand{\Q}{\ensuremath{\text{\textsf{Q}}}\xspace}
 
 \renewcommand{\S}{\ensuremath{\text{\textsf{S}}}\xspace}
 \renewcommand{\H}{\ensuremath{\text{\textsf{H}}}\xspace}
 
 \newcommand{\Face}{\ensuremath{\text{\textsl{Face}}}\xspace}
 \newcommand{\Suit}{\ensuremath{\text{\textsl{Suit}}}\xspace}
 
 \newcommand{\Obs}{\ensuremath{\mathcal{V}}\xspace}

\begin{document}
 \makeatletter
 \def\ps@titlepage{%
   \renewcommand{\@oddfoot}{}%
   \renewcommand{\@evenfoot}{}%
   \renewcommand{\@oddhead}{\hfill\arXiv}
   \renewcommand{\@evenhead}{}}
 \makeatother

%%%%%%%%%%%%%%%%%%%%%%%%%%%%%%%%%%%%%%%%%%%%%%%%%%%%%%%%%%%%%%%%%%%%%%%%%%%%%%%%%%%%%%%
\title[Kirkpatrick -- \Title] %% for running titles on pages
      {\Title} %% the title-page title

%%%%%% PERSONAL %%%%%%%%%%%%%%%%%%%%%%%%%%%%%%%%%%%%%%%%%%%%%%%%%%%%%%%%%%%%%%%%%%%%%%%%
\author{K.~A.~Kirkpatrick}
\email[E-mail: ]{kirkpatrick@physics.nmhu.edu}
%\homepage[]{}
\affiliation{New Mexico Highlands University, Las Vegas, New Mexico 87701}
%%%%%%%%%%%%%%%%%%%%%%%%%%%%%%%%%%%%%%%%%%%%%%%%%%%%%%%%%%%%%%%%%%%%%%%%%%%%%%%%%%%%%%%%
%%%%%% ABSTRACT %%%%%%%%%%%%%%%%%%%%%%%%%%%%%%%%%%%%%%%%%%%%%%%%%%%%%%%%%%%%%%%%%%%%%%%%
\begin{abstract}
 \Abstract
\end{abstract}
% insert suggested PACS numbers in braces on next line
% \pacs{03.65.Bz, 01.70.+w}

%%%%%%%%%%%%%%%%%%%%%%%%%%%%%%%%%%%%%%%%%%%%%%%%%%%%%%%%%%%%%%%%%%%%%%%%%%%%%%%%%%%%%%%%
 \maketitle
%%%%%%%%%%%%%%%%%%%%%%%%%%%%%%%%%%%%%%%%%%%%%%%%%%%%%%%%%%%%%%%%%%%%%%%%%%%%%%%%%%%%%%%%
%% centered short title in each header:
 \makeatletter\markboth{\hfill\@shorttitle\hfill}{\hfill\@shorttitle\hfill}\makeatother
 \pagestyle{myheadings}
%%%%%%%%%%%%%%%%%%%%%%%%%%%%%%%%%%%%%%%%%%%%%%%%%%%%%%%%%%%%%%%%%%%%%%%%%%%%%%%%%%%%%%%%
%%% BODY OF DOCUMENT %%%%%%%%%%%%%%%%%%%%%%%%%%%%%%%%%%%%%%%%%%%%%%%%%%%%%%%%%%%%%%%%%%%
%%%%%%%%%%%%%%%%%%%%%%%%%%%%%%%%%%%%%%%%%%%%%%%%%%%%%%%%%%%%%%%%%%%%%%%%%%%%%%%%%%%%%%%

Hardy's Axiom 2 states\cite{Hardy01} that the informational degrees of freedom of a
system, $K$, is a function of the number of values a variable is allowed, $N$: $K=K(N)$.
He analyzes a classical probability system (balls in urns) to show that, in classical
probability, $K=N$; in quantum mechanics, $K=N^2$. Hardy suggests that this difference
characteristically distinguishes classical from quantal.

This is contradicted by the following simple example from classical probability (of a
type I've discussed in \RefCites{Kirkpatrick:Quantal} and
\onlinecite{Kirkpatrick:ThreeBox}):
\begin{quote}
The system has $V$ discrete-valued variables, call them \Face, \Suit, \dots, each taking
on $N$ values, respectively \set{\K,\,\Q,\,\cdots}, \set{\S,\,\H,\,\cdots}, \dots. It
consists of a box with a single display and pushbutton (momentary) switches marked
``\Face,'' ``\Suit,'' \dots. Inside the box is a deck of playing cards, each marked with
a \Face--value, a \Suit--value, \dots (the cards are not necessarily unique). Some of the
cards are segregated into a subdeck.

When the switch labeled \Obs\ (\ie \Face, \Suit, \dots) is pressed, a card is selected
from the subdeck by a nondeterministic process, and the value $y$ of the card's variable
\Obs\ is shown momentarily in the display. A new subdeck is constructed, consisting of
all the cards of the deck for which the value of the variable \Obs\ is $y$.
%
%The system may be prepared in a state sharp in any variable's value: the preparation
%apparatus presses the appropriate switch and then filters the systems, by repeatable
%observation, on the desired value. (This has the effect of creating a subdeck containing
%all the cards having the desired value.)
\end{quote}
In this system occurrences (``observations'') are repeatable: If, for a given occurrence,
$\Suit=\H$ say, then a succeeding observation of \Suit\ will return $\Suit=\H$ with
certainty, while a succeeding observation of \Face\ will randomly return \K\ or \Q\ or
\dots. (These variables are incompatible: they do not have a joint probability
distribution, nor are they simultaneously observable.)

It is necessary to determine the probability of each value of each variable in order to
fully determine the state; thus, in this example, $K=NV$. Of course if $V=1$ (as is the
case for his classical example, a ball in boxes), this corresponds to Hardy's
``classical'' case, $K=N$. The maximal number of independent variables is equal to the
dimension of the system: $V=N$, in which case $K=N^2$. This is the relation Hardy
obtained for the quantum case---but this example is certainly not quantal.

In a quantum mechanical system of $N$ dimensions, there are maximally $V=N$ linearly
independent variable operators, so $K=N^2$. However, if superselection rules are active,
then there are fewer physical variables, $V<N$. Similarly, in the classical example
presented here, the number of variables needn't be maximal, so again it is possible that
$V<N$.

Both in classical systems of the form exemplified here and in quantal systems we have
$K=VN$, $1<V<N$. The functional form of $K$ cannot distinguish classical probabilistic
systems from quantal systems.

\enlargethispage*{1em}%%LO
%%%%%%%%%%%%%%%%%%%%%%%%%%%%%%%%%%%%%%%%%%%%%%%%%%%%%%%%%%%%%%%%%%%%%%%%%%%%%%%%%%%%%%%%
%%% END OF BODY %%%%%%%%%%%%%%%%%%%%%%%%%%%%%%%%%%%%%%%%%%%%%%%%%%%%%%%%%%%%%%%%%%%%%%%%
%%%%%%%%%%%%%%%%%%%%%%%%%%%%%%%%%%%%%%%%%%%%%%%%%%%%%%%%%%%%%%%%%%%%%%%%%%%%%%%%%%%%%%%%

%%%%%%%%%%%%%%%%%%%%%%%%%%%%%%%%%%%%%%%%%%%%%%%%%%%%%%%%%%%%%%%%%%%%%%%%%%%%%%%%%%%%%%%%
%% BIBLIOGRAPHY %%%%%%%%%%%%%%%%%%%%%%%%%%%%%%%%%%%%%%%%%%%%%%%%%%%%%%%%%%%%%%%%%%%%%%%%
%%%%%%%%%%%%%%%%%%%%%%%%%%%%%%%%%%%%%%%%%%%%%%%%%%%%%%%%%%%%%%%%%%%%%%%%%%%%%%%%%%%%%%%%
 \renewcommand{\refname}{\sc References}%xxx
 \footnotesize%
%%%%%%%%%%%%%%%%%%%%%%%%%%%%%%%%%%%%%%%%%%%%%%%%%%%%%%%%%%%%%%%%%%%%%%%%%%%%%%%%%%%%%%%
%% \bibliographystyle{myapsrev}
% \bibliographystyle{ajp}
%% \bibliographystyle{myapalike}
% \bibliography{JAbbrevs,QM\NotesBib}

%% For submission, comment out previous lines, follow instructions at "Footnotes in
%% Bibliography," above, and copy .bbl here:

%%%%%%%%%%%%%%%%%%%%%%%%%%%%%%%%%%%%%%%%%%%%%%%%%%%%%%%%%%%%%%%%%%%%%%%%%%%%%%%%%%%%%%%%
%%% END DOCUMENT %%%%%%%%%%%%%%%%%%%%%%%%%%%%%%%%%%%%%%%%%%%%%%%%%%%%%%%%%%%%%%%%%%%%%%%
%%%%%%%%%%%%%%%%%%%%%%%%%%%%%%%%%%%%%%%%%%%%%%%%%%%%%%%%%%%%%%%%%%%%%%%%%%%%%%%%%%%%%%%%
\end{document}